\documentclass[letter,bibyear]{aa} 

\usepackage{epsfig}
\usepackage{amsmath}
\usepackage{subfigure}
\usepackage{natbib}
\usepackage{multicol}
\usepackage{multirow}
\usepackage{color}
\usepackage{float}
\usepackage{longtable}
\usepackage{graphicx}
\usepackage{epstopdf}
\usepackage{booktabs}

\usepackage{stackrel,amssymb,amsmath}
\newcommand{\leftrarrows}{\mathrel{\raise.75ex\hbox{\oalign{%
  $\scriptstyle\leftarrow$\cr
  \vrule width0pt height.5ex$\hfil\scriptstyle\relbar$\cr}}}}
\newcommand{\lrightarrows}{\mathrel{\raise.75ex\hbox{\oalign{%
  $\scriptstyle\relbar$\hfil\cr
  $\scriptstyle\vrule width0pt height.5ex\smash\rightarrow$\cr}}}}
\newcommand{\Rrelbar}{\mathrel{\raise.75ex\hbox{\oalign{%
  $\scriptstyle\relbar$\cr
  \vrule width0pt height.5ex$\scriptstyle\relbar$}}}}

\usepackage{txfonts}

\newcommand{\jmst}{J.~Mol.~Struct.}   

\begin{document}

\title{The spatial distribution of an aromatic molecule, C$_6$H$_5$CN, in the cold dark cloud TMC-1\thanks{Based on observations carried out with the Yebes\,40m telescope
(projects 22A007 and 22B029).
The 40m radio telescope at Yebes Observatory is operated by the Spanish Geographic 
Institute (IGN, Ministerio de Fomento). }}

\author{
J.~Cernicharo\inst{1},
B.~Tercero\inst{2,3},
N.~Marcelino\inst{2,3},
M.~Ag\'undez\inst{1},
and
P.~de~Vicente\inst{2}
}
\institute{Dept. of Molecular Astrophysics, Instituto de F\'isica Fundamental (IFF-CSIC),
C/ Serrano 121, 28006 Madrid, Spain\\ \email jose.cernicharo@csic.es
\and Centro de Desarrollos Tecnol\'ogicos, Observatorio de Yebes (IGN), 19141 Yebes, Guadalajara, Spain
\and Observatorio Astron\'omico Nacional (OAN, IGN), Madrid, Spain
}

\date{Received 21 April 2023; accepted 21 May 2023}

\abstract{
We present a highly sensitive 2D line survey of TMC-1 obtained with the Yebes 40m
radio telescope in the Q-band (31.13-49.53 GHz). These maps cover a region of 
320$''$$\times$320$''$ centred
on the position of the QUIJOTE$^1$ line survey with a spatial sampling of 20$''$. The 
region covering 
240$''$$\times$240$''$, where a longer integration time
was used, shows a homogenous sensitivity of 2-4\,mK across the band. 
We present in this work the first determination of the spatial extent of benzonitrile
(C$_6$H$_5$CN), which follows  that of cyanopolyynes rather well, but differs
significantly from
that of the radicals C$_n$H and C$_n$N. We definitively conclude that aromatic species
in TMC-1 are formed from chemical reactions involving smaller species
in the densest zones of the cloud.
}


\titlerunning{C$_6$H$_5$CN in TMC-1}
\authorrunning{Cernicharo et al.}

\maketitle

\section{Introduction}
In the last five years, two line surveys of the starless cold core TMC-1, 
QUIJOTE\footnote{\textbf{Q}-band \textbf{U}ltrasensitive \textbf{I}nspection \textbf{J}ourney
to the \textbf{O}bscure \textbf{T}MC-1 \textbf{E}nvironment} 
\citep{Cernicharo2020,Cernicharo2021a,Cernicharo2023}
and GOTHAM\footnote{
\textbf{G}BT \textbf{O}bservations of \textbf{T}MC-1: \textbf{H}unting for \textbf{A}romatic \textbf{M}olecules}
\citep{McGuire2018, McGuire2020}, have been carried out and have 
provided the discovery of a panoply of new radicals, aromatic and polyaromatic compounds, cations, 
anions, and sulphur-bearing species. Roughly one-third of the molecules discovered in space have resulted from these two line surveys.
The GOTHAM observations were taken with the Green Bank 100m radio telescope in the X-, K-, and Ka-bands.
Most detections in this line survey were performed through a sophisticated statistical
frequency stacking procedure.
The QUIJOTE line survey is an ongoing line survey in the Q-band (31.1-50.3 GHz) obtained with the Yebes 40m radio
telescope. The detection technique in QUIJOTE is the classical, and reliable, line-by-line detection without
any spectral stacking. The present sensitivity of QUIJOTE is 0.12-0.25 mK \citep{Cernicharo2023}.
 
The modelling
of this emission is often tackled with  very limited information on the spatial
extent of the observed emission. In GOTHAM and QUIJOTE the only available spatial information
is provided by the variation of the telescope  half power beam with the frequency across the
line survey. While GOTHAM
fits four velocity components with different spatial sizes, QUIJOTE assumes a source radius
of 40$''$ based on previous observations of TMC-1 in several molecular species. None of these
methods is satisfactory to obtain the  accurate column densities needed to put constraints on the  chemical models of the source. 
The spatial size of the observed molecules, together 
with the issues related to the line opacities and radiative transfer, can only be 
addressed through spatial mapping of 
the molecular emission. 

To overcome these issues, the 
QUIJOTE line survey has been complemented with high-sensitivity maps obtained
with the Yebes 40m radio telescope and covering
a region of 320$''$$\times$320$''$  centred
on the QUIJOTE position (see Sect. \ref{observations}). 
These maps are a faithful companion to the QUIJOTE line survey and
we  call these 
supplementary spatial data SANCHO.\footnote{
\textbf{S}urveying the \textbf{A}rea of the \textbf{N}eighbour TMC-1 \textbf{C}loud 
through \textbf{H}eterodyne \textbf{O}bservations} The goal of these maps is to 
permit the study of
the spatial distribution 
of any QUIJOTE line with intensity $\geq$20 mK with a signal-to-noise ratio $\geq$10, which 
means the low-energy transitions of most of the abundant species such as 
cyanopolyynes and radicals (C$_n$S, C$_n$H, and C$_n$N)  among other molecules.
Most of the isotopologues 
$^{13}$C, $^{34}$S, D, and $^{15}$N of these species can be also
spatially traced with SANCHO data.

In this letter we present the first determination of the spatial distribution of an aromatic molecule
(benzonitrile, C$_6$H$_5$CN) in the starless cold dark cloud TMC-1, and compare it with 
that of other molecules. 

\begin{figure*}
\centering
\includegraphics[width=1\textwidth,angle=0]{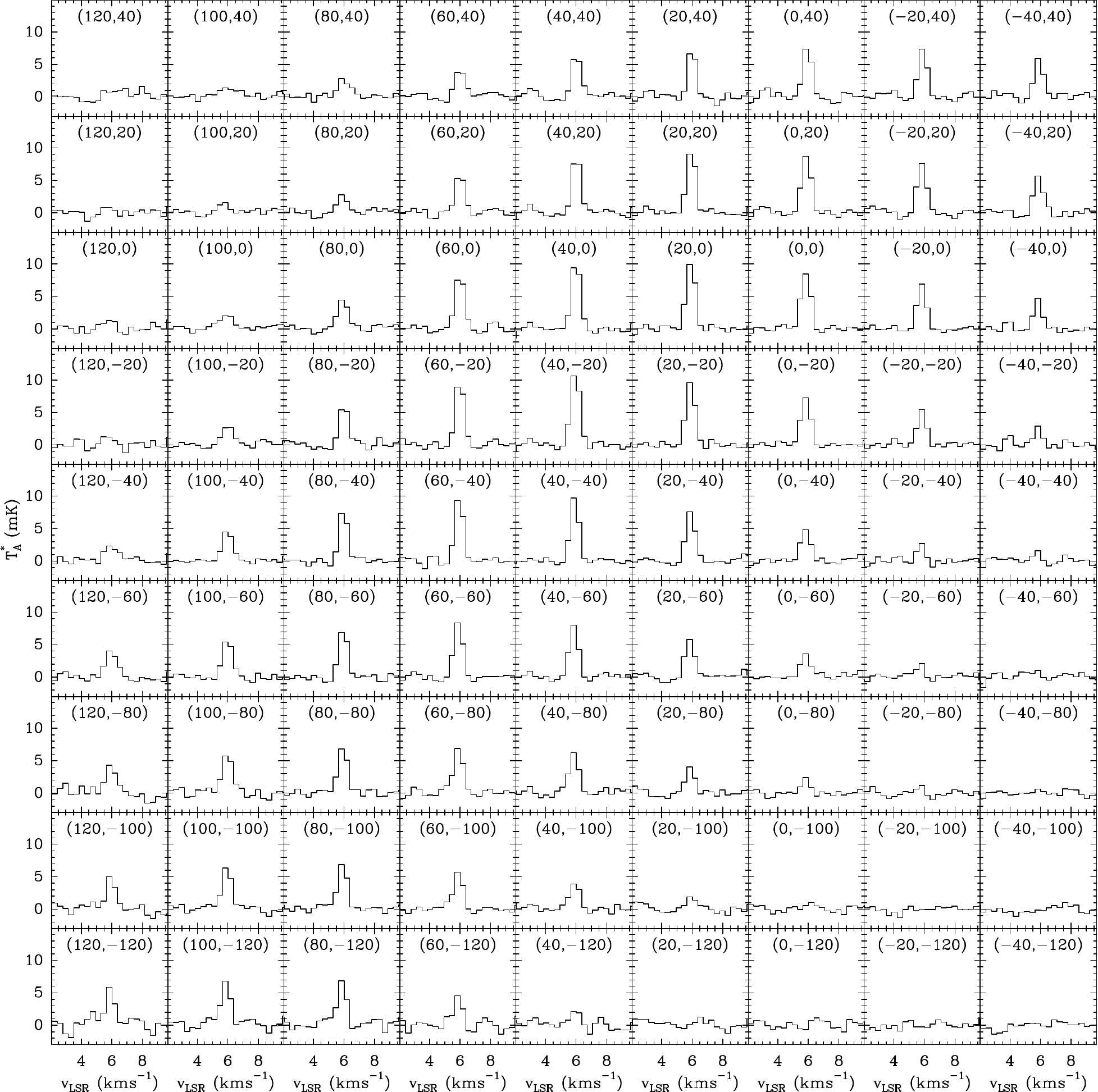}
\caption{Stacked line of C$_6$H$_5$CN obtained from the average in each
position of 52 individual lines (see Appendix \ref{app:stacked}). Only  the spectra around the region of maximum emission (see Fig. \ref{fig:area}) are shown. 
The offset positions in
arcseconds are indicated in each panel. The abscissa in each panel corresponds to the local standard of rest 
velocity of the stacked data (in km\,s$^{-1}$).
The intensity scale corresponds to the antenna temperature corrected for atmospheric transmission and antenna losses.}
\label{fig:spectra}
\end{figure*}

\section{Observations} \label{observations}
New receivers, built within the Nanocosmos\footnote{ERC grant ERC-2013-Syg-610256-NANOCOSMOS.\\
https://nanocosmos.iff.csic.es/} project
and installed at the Yebes 40m radiotelescope, were used
for the observations of TMC-1
($\alpha_{J2000}=4^{\rm h} 41^{\rm  m} 41.9^{\rm s}$ and $\delta_{J2000}=
+25^\circ 41' 27.0''$). 
The present observations of TMC-1 complement the QUIJOTE line survey 
\citep{Cernicharo2021a}. A detailed description of the telescope, receivers, and backends is 
given in \citet{Tercero2021}. Briefly, the receiver consists of two cold high electron 
mobility transistor amplifiers covering the 
31.1-50.3 GHz band with horizontal and vertical             
polarisations. Receiver temperatures in the first QUIJOTE observations achieved during 2020 
vary from 22 K at 32 GHz to 42 K at 50 GHz. However, some power adaptation in the down-conversion 
chains in 2021 reduced
the receiver temperatures to 16\,K at 32 GHz and 25\,K at 50 GHz.
The backends are $2\times8\times2.5$ GHz fast Fourier transform spectrometers (FFTs)
with a spectral resolution of 38.15 kHz
providing the whole coverage of the Q-band in both polarisations.
The final Doppler correction was made during the pipeline processing of 
the raw data and the final spectra have the correct frequency scale in every part 
of the 31.1-50.3 GHz band.

The telescope beam size varies from 56.7$''$ at 31 GHz to 35.6$''$ at 49.5 GHz.
The intensity scale used in this work, antenna temperature
($T_A^*$), was calibrated using two absorbers at different temperatures and the
atmospheric transmission model (ATM) \citep{Cernicharo1985, Pardo2001}.
Calibration uncertainties of 10~\% were adopted.  
The beam efficiency of the Yebes 40m telescope in the Q-band is given as
a function of frequency by $B_{\rm eff}$=0.797 exp[$-$($\nu$(GHz)/71.1)$^2$]. The
forward telescope efficiency is 0.95.

The SANCHO maps were performed in the on-the-fly mode using
frequency switching with a throw of 10 MHz. We prefer this observing mode as
many molecular species have extended emission \citep{Cernicharo1987}.
The speed of the telescope was 5$''$/sec
and the data of the whole Q-band in the two polarisations were recorded every two seconds (16 individual spectra per position).
The maps were done moving the telescope in right ascension from $-$130$''$ to +130$''$ with a declination step of 10$''$. 
Once such a map was achieved, then the telescope was moved in declination from $-$130$''$ to 130$''$ with a
right ascension step of 10$''$. This procedure was repeated until we reached 100 hours
of observing time on the source. 
Some additional maps with a total observing time on source of 15 hours
were added, covering offsets up to $\pm$160$''$ in right ascension and declination. 
A total of 3468624 spectra, corresponding to 216789 different positions, two polarisations, and eight FFTs, were 
recorded in February and December 2022, and January 2023. The raw data file size is 
near 1 TByte. All data were analysed using the GILDAS package.
\footnote{\texttt{http://www.iram.fr/IRAMFR/GILDAS}}

\section{Results} \label{results}

For the final SANCHO data, we produced six different types of maps that were obtained by resampling the raw data within a square grid with a point separation 
of 10$''$ and 20$''$ ($S_{grid}$), and with three different spatial tolerances around each position for adding raw data. 
These tolerance zones correspond to circles of 10$''$, 15$''$, and 20$''$ in radius ($T_{rad}$). Each position in the maps 
contains the whole Q-band spectrum, with both polarisations averaged. The highest sensitivity is achieved  for the maps with $S_{grid}$\,=\,20$''$ and $T_{rad}$\,=\,20$''$, which are those that 
we discuss in this work (see Appendix \ref{sampling}). In this process no baseline is applied as the whole Q-band spectrum 
is treated simultaneously (31.13-49.53 GHz with
483928 channels at each position of the map). These maps are fully sampled spatially.
However, due to the spatial tolerance used in creating the maps we expect to have produced some
spatial smoothing (i.e. the emission in adjacent positions has some degree of correlation; 
see Appendix \ref{sampling}).

\begin{figure}
\centering
\includegraphics[width=0.49\textwidth,angle=0]{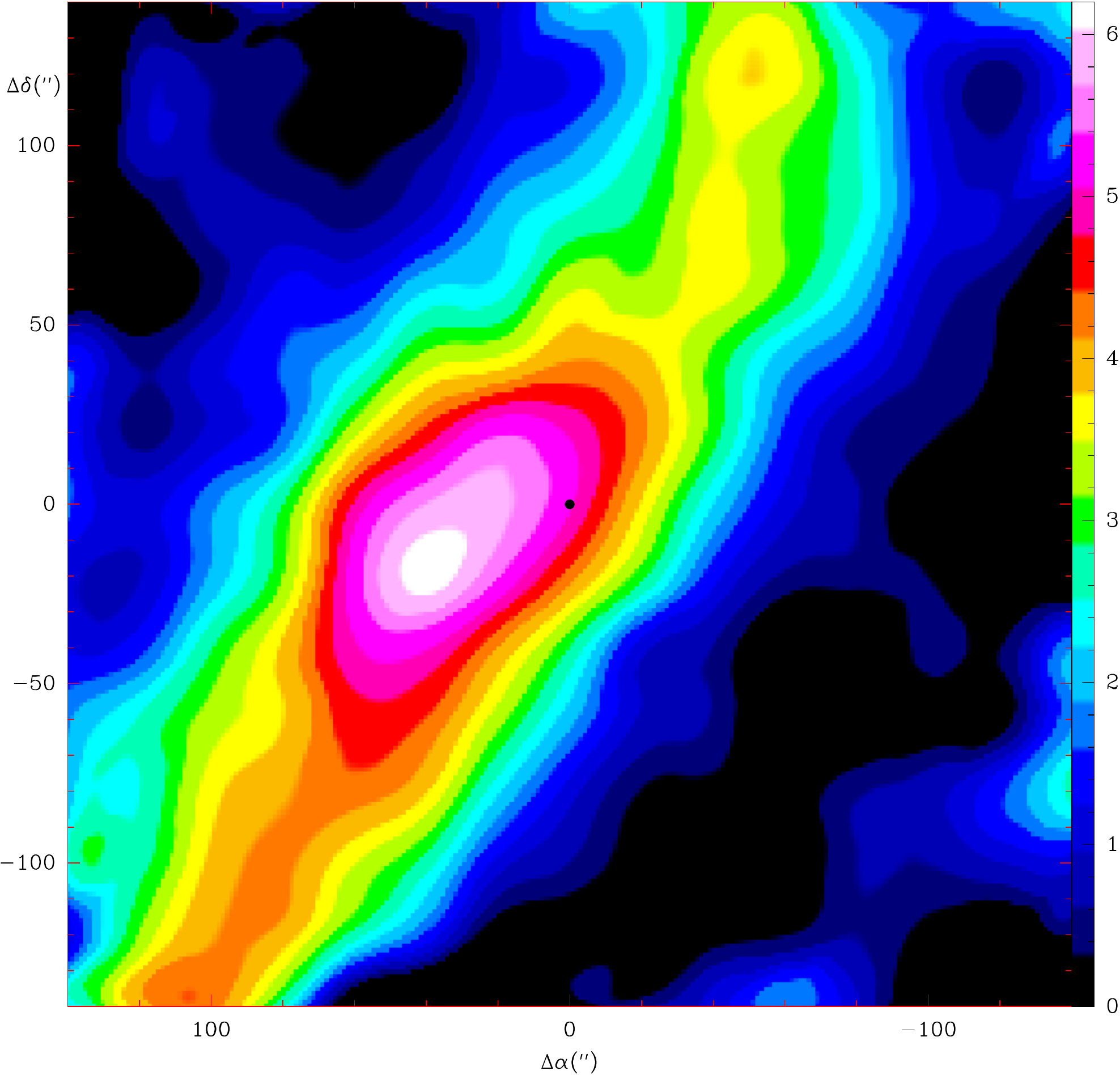}
\caption{Colour plot of  spatial distribution of the integrated intensity between 5.3 and 6.5 km\,s$^{-1}$
of the stacked lines of C$_6$H$_5$CN.
The colour scale is in mK km\,s$^{-1}$. The sampling of the data is 20$''$ and the
integrating area in each position corresponds to a circle of 20$''$ of radius.
The black dot indicates the centre of the map, which corresponds to the
position observed with the QUIJOTE line survey.}
\label{fig:area}
\end{figure}

Maps for each individual transition of a given molecule were obtained from the 
selected final map (with a given $S_{grid}$ and $T_{rad}$) by extracting
the spectral data over $\pm$14 km\,s$^{-1}$ around the frequency of the line.
Transition frequencies were obtained
from the spectral information of the MADEX code \citep{Cernicharo2012}, the JPL catalogue \citep{Pickett1998}, or the CDMS database \citep{Muller2005}.
A baseline is removed from each extracted spectrum after defining the windows for all lines potentially
present in the selected velocity range using the QUIJOTE line survey as a reference for
detected lines (only features above 5 mK were blanked in this process). 
Appendix \ref{sampling} provides examples of several molecular lines at the central position
of the map without and with baseline removal. 
The sensitivity over the Q-band as a function of frequency, $T_{rad}$, 
and position in the map is provided in Appendix 
\ref{sampling} (see Table \ref{tab:sensitivity}).

It is worth noting that all maps produced in this way were observed simultaneously, and hence
they have the same pointing and the same calibration uncertainties. Moreover, the relative calibration between different lines of a given molecule
is much better than the global calibration uncertainty of 10\,\%  (see e.g. 
Appendix \ref{sampling} and Fig. \ref{fig:c3n_lines}).
The achieved sensitivity in each position depends on the adopted $T_{rad}$ in adding
data. In Appendix \ref{sampling} we show the effect of adopting different gridding
steps and tolerance circles. For $S_{grid}$\,=\,20$''$ and $T_{rad}$\,=\,20$''$, the
sensitivity is 1.4-4 mK across the Q-band (the sensitivity is the same for $S_{grid}$=10$''$). This sensitivity is 
degraded by a factor of $\sim$2 outside the square defined by $\Delta\alpha$=$\pm$120$''$ and 
$\Delta\delta$=$\pm$120$''$.
Hence, SANCHO provides a whole spectrum of the Q-band over a map
of 169 positions gridded every 20$''$ (240$''$$\times$240$''$) with a sensitivity that 
is similar to that achieved by GOTHAM in a single
position. However, the sensitivity towards the centre of the map is worse by a factor of 
10 compared to that of the QUIJOTE line survey.

\begin{figure*}
\centering
\includegraphics[width=1\textwidth,angle=0]{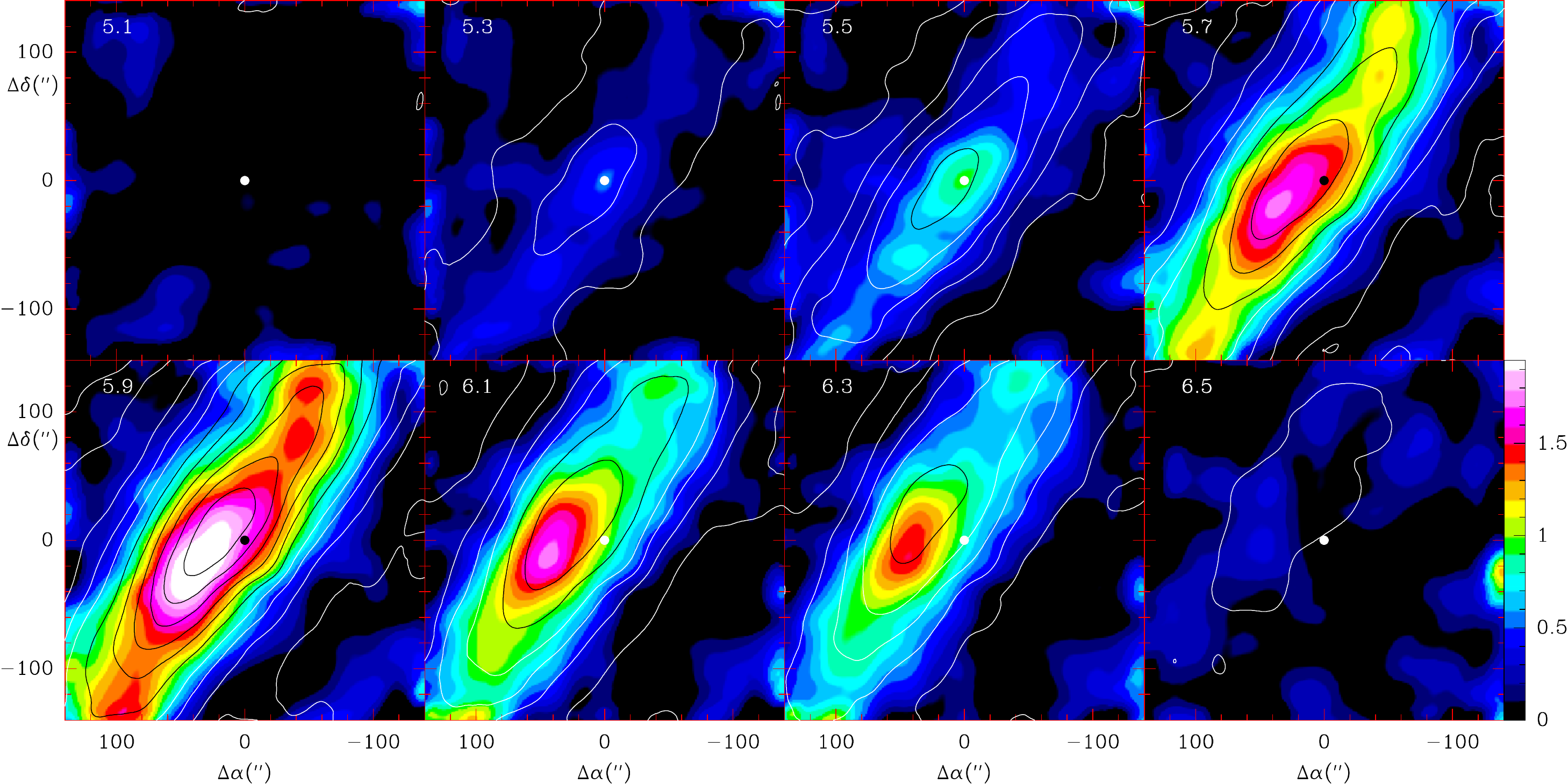}
\caption{Spatial distribution of the integrated intensity of the stacked lines of C$_6$H$_5$CN at different velocities. The selected 
velocity range is $\pm$0.1 km\,s$^{-1}$ around the velocity indicated
in the upper right corner of each panel (in km s$^{-1}$). The sampling of the data is 20$''$ and the
integrating area in each position corresponds to a circle of 20$''$  radius.
The white and black contours represent the integrated intensity of the $J$=28-27 transition of
HC$_7$N at the same velocity ($\pm$0.1 km\,s$^{-1}$).
The first white contour corresponds to 2 mK kms$^{-1}$ and the step is 4 mK kms$^{-1}$. The first
black contour corresponds to 20 mK kms$^{-1}$ and the step is 6 mK kms$^{-1}$. The white or black dot indicates
the centre of the map.}
\label{fig:vel}
\end{figure*}


The sensitivity reached by SANCHO allows the  detection of  all lines with intensities greater than 10 mK in the central position
of the maps. This is enough to detect many of the lines of C$_6$H$_5$CN observed with the QUIJOTE line survey
presented in Fig. B.1 of \cite{Cernicharo2021b}. However, the spatial distribution of the integrated intensity
of each individual line appears noisy, which prevents us from reaching a conclusion on the spatial extent of this 
aromatic species. We therefore used the maps generated for each individual line to produce a spectrally stacked map. We added 49 lines of benzonitrile which are
free of resolved hyperfine structure and of blending with lines from other species (see Table \ref{tab:c6h5cn_lines}). 
Each line was multiplied by a factor to take into account the observed intensity with respect 
to a reference line, the strongest one of benzonitrile
in the QUIJOTE line survey (the 13$_{0,13}$-12$_{0,12}$ transition at 32833.827 MHz).
The procedure is described in detail in Appendix \ref{app:stacked}.
The stacking scheme allows one to use the parameters of the reference line to estimate 
the column density and increases the S/N of the stacked spectral map 
with respect the observed individual line maps of benzonitrile.

The resulting spectra at different
positions in the map are shown in Fig. \ref{fig:spectra}. The sensitivity of the map of the stacked lines at the ($\Delta\alpha$, $\Delta\delta$) 
positions (0$''$, 0$''$), (120$''$, 0$''$), and (120$''$, $-$120$''$) is 0.4 mK, 0.6 mK, and 0.9 mK, respectively (see  Fig. \ref{fig:spectra}).
The spatial distribution of the integrated intensity of this aromatic species is shown in Fig. \ref{fig:area}. Finally, the spatial
distribution of the emission at different velocities is shown in Fig. \ref{fig:vel}.

\begin{figure*}
\centering
\includegraphics[width=1\textwidth,angle=0]{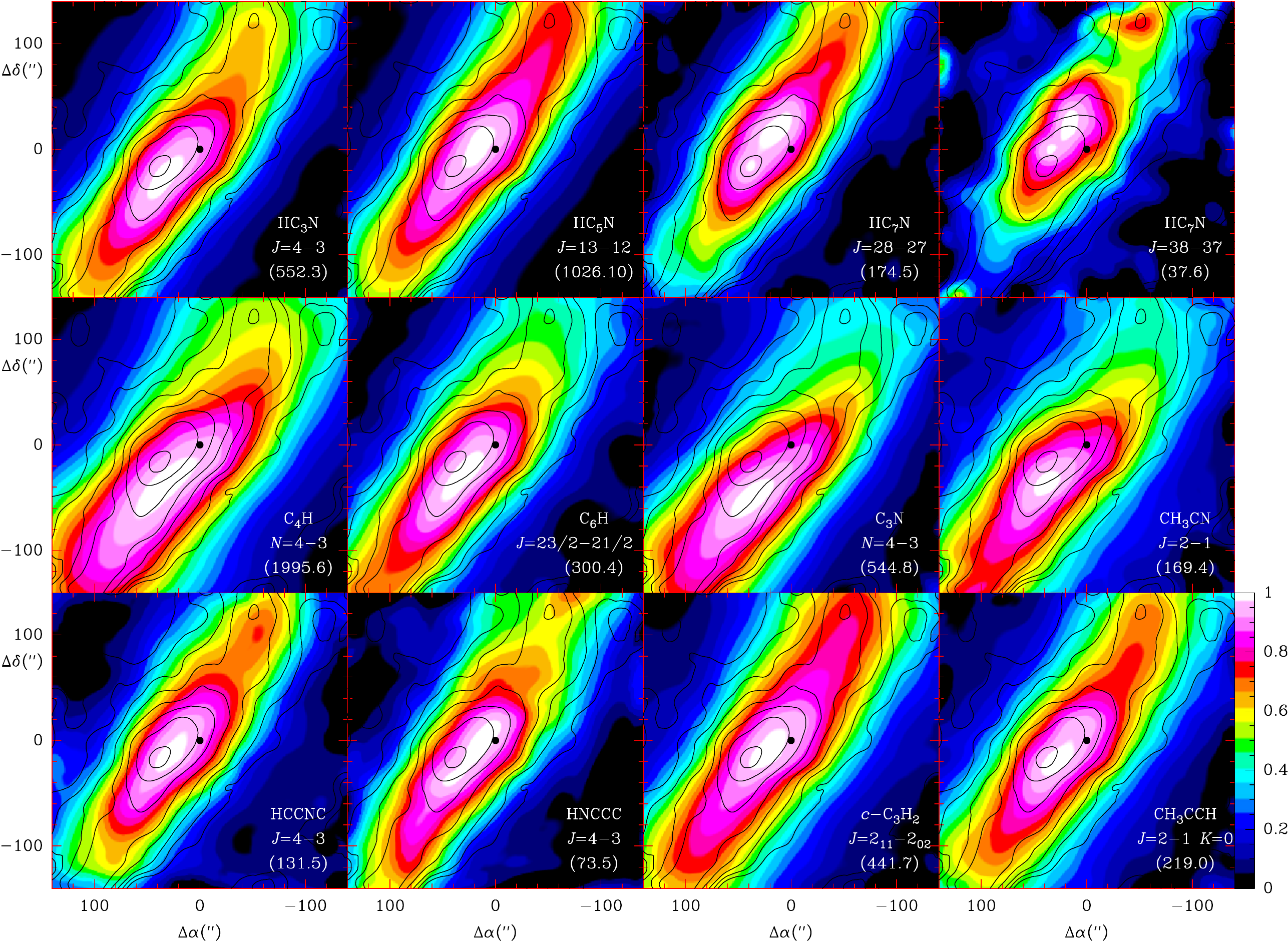}
\caption{Integrated intensity between 5.3 and 6.5 kms$^{-1}$
of different molecular transitions (colours) compared with that of
C$_6$H$_5$CN (black contours; first contour and step are 0.75 mK kms$^{-1}$). For each molecular
transition the integrated intensity has been normalised to the maximum value
within the area covered by the map. The colour scale (bottom right) is
the same for all molecular transitions. The molecule, transition, and maximum intensity (shown in brackets) 
are indicated in the bottom right corner of each panel.
The black dot corresponds to the centre of the map.}
\label{fig:compa}
\end{figure*}

\section{Discussion} \label{discussion}

The comparison of the velocity maps of C$_6$H$_5$CN and HC$_7$N shown in Fig. \ref{fig:vel} clearly  indicates
that the two species coexist spatially. The emission of the two species covers the velocity range 5.3-6.3 km\,s$^{-1}$, with
some weak emission still at 6.5 km\,s$^{-1}$. The observed emission follows the well-known 
filamentary structure of TMC-1 with
a dense condensation around the central position. Two different velocities seem to be traced 
spatially by our data, 
both  with a filamentary aspect: one
at 5.5 km\,s$^{-1}$ with its maximum over the QUIJOTE position (known as the  cyanopolyynes source) 
and the other   at 
5.9 km\,s$^{-1}$ that peaks at $\Delta\alpha$=40$''$ and $\Delta\delta$=-20$''$, which dominates the 
line profile. Given our velocity resolution of $\sim$0.25 km\,s$^{-1}$, the
observed velocity structure could also be interpreted as resulting from a velocity gradient
perpendicular to the TMC-1 filament, or as a twisting motion of filaments, or fibres, similar 
to those found in other molecular clouds \citep{Kirk2013,Hacar2013,Hacar2017}.
This behaviour is
similar for all the species we  study in this work. The TMC-1 filament was 
observed in the emission of molecules such as SO, NH$_3$, and CCS, with higher spatial resolution 
than we used 
\cite[see e.g.][and references therein]{Lique2006,Feher2016,Dobashi2019}. 
These authors found a complex spatial velocity 
structure dominated by clumpy filaments at velocities between 5.3 and 6.3 km\,s$^{-1}$.

The spatial coexistence of benzonitrile and other molecules is fully confirmed 
in Fig. \ref{fig:compa}, where
we compare the spatial distribution of
the integrated intensity of several species, including radicals, cyanopolyynes, methyl-bearing species, and isomers
of HC$_3$N. This is the
first time that many of the selected molecules have been mapped in TMC-1. Moreover, the signal-to-noise ratio of
all these maps is extremely high as the selected lines have intensities $\geq$100 mK, even for the HC$_7$N $J$=38-37
for which the upper level energy is 40.1\,K. 
The integrated intensities are proportional to the column 
densities because the lines are mostly optically thin. The map of HC$_3$N was obtained
using the integrated intensity of the weak satellite hyperfine lines, which also guarantees 
low line opacities.

Opacity effects, the beam size of the telescope, different excitation conditions, 
and the spatial smearing produced by the co-addition of the on-the-fly data
could produce some small differences between the maps. However,
some of these differences are real and are related to the global aspect of the spatial distribution. For example,
cyanopolyynes and benzonitrile  are significantly  shifted towards the north-east by more
than a beam with respect to the radicals C$_n$H and C$_n$N. 
The spatial shift of the radicals C$_4$H, C$_6$H, and C$_3$N with respect to the 
cyanopolyynes HC$_3$N and HC$_5$N may be caused by the structure of the H$_2$ density. 
The HC$_3$N peak is also a local maximum in H$_2$ density \cite[see][]{Pratap1997}, 
and radicals are probably more efficiently destroyed at high densities than closed-shell molecules.
A different spatial distribution
was also found in TMC-1 by \cite{Fosse2001} for C$_6$H, $c$-C$_3$H$_2$, and H$_2$C$_3$.
Significant differences are also found for the
two transitions of HC$_7$N shown in Fig. \ref{fig:compa}. This is probably due to the high energy of 
the $J$=38 level,
which renders this transition very sensitive to the density or to small changes in the kinetic 
temperature. The analysis of the spatial distribution of several transitions of radicals, 
cyanopolyynes, hydrocarbons, and their isotopologues will be published in forthcoming papers.

The formation of aromatic cycles in TMC-1 has been a matter of debate in recent years 
\citep{McGuire2018,McGuire2021,Cernicharo2021b,Cernicharo2022}. It is not yet clear whether these aromatic 
rings are formed through a bottom-up mechanism or a top-down process. The presence of large polycyclic aromatic hydrocarbons (PAHs) in 
diffuse interstellar clouds is inferred from the observation of intense unidentified infrared bands, and the 
aromatic cycles observed in TMC-1 could result from the destruction of this reservoir of PAHs inherited from 
a previous diffuse stage. A similar process has been invoked to account for the formation of small hydrocarbons 
in photodissociation regions \citep{Pety2005}. Alternatively, aromatic rings could be formed from small hydrocarbons 
in a bottom-up process. For example, the reaction of propene, an abundant hydrocarbon in TMC-1 \citep{Marcelino2007}, 
with CH can produce 1,3-butadiene \citep{Loison2009}, which can lead to $c$-C$_5$H$_7$ upon reaction with CH 
\citep{Cernicharo2021b} and to benzene by reacting with C$_2$H \citep{Jones2011}. Even so, chemical models based 
on these neutral-neutral routes are not able to account for the observed abundances of cycles such as 
$c$-C$_5$H$_7$ \citep{Cernicharo2022}, which indicates that ion-neutral reactions may also play a role 
in the synthesis of aromatic cycles. Whatever  the true mechanism of formation of aromatic rings, 
the fact that benzonitrile has a spatial distribution very similar to that of cyanopolyynes supports the argument 
that aromatic molecules are formed through a bottom-up chemical process, as  is thought to occur for 
long carbon chains in TMC-1.

\section{Conclusions}

We have presented the SANCHO maps of TMC-1 in the frequency range 31.13-49.53 GHz. The sensitivity of the
spectra over an area of 240$''$$\times$240$''$, for a gridding step of 20$''$ and $T_{rad}$=20$''$, 
is 2-4 mK across the band. We have determined, for the first time, the spatial distribution of
C$_6$H$_5$CN (benzonitrile). We find that it correlates very well with that of cyanopolyynes, and that
it extends over the TMC-1 filament. Hence, we conclude that this species is formed through bottom-up
chemical processes involving the same type of reactions as those forming the other
species found in the cloud.

\begin{acknowledgements}

We thank ERC for funding
through grant ERC-2013-Syg-610256-NANOCOSMOS. We also thank Ministerio de Ciencia e Innovaci\'on of Spain (MICIU) for funding support through projects
PID2019-106110GB-I00, PID2019-107115GB-C21 / AEI / 10.13039/501100011033, 
and PID2019-106235GB-I00. 

\end{acknowledgements}

\begin{appendix}
\section{Baselines, gridding step, tolerance circle, and sensitivity}
\label{sampling}
Our on-the-fly frequency switching observations suffer from frequency ripples. However, the period of these ripples
is always larger than 10 MHz, while the lines in the Q-band are typically 0.10-0.18 MHz wide. Hence, 
it is possible to remove a
polynomial baseline to each line of the survey at each position of the map with total confidence 
that neither the line intensity nor the line profile are perturbed by the baseline removal step.
These effects are also present in the QUIJOTE line survey, and we follow a procedure similar
to that previously described for these frequency switching observations \citep{Cernicharo2022}.
\begin{table}
\centering
\caption{Sensitivity of the maps as a function of the frequency
and of the distance to the central position.}
\label{tab:sensitivity}
\begin{tabular}{|c|c|c|c|}
\hline
Frequency  & $\sigma(10$''$)^a$ & $\sigma(15$''$)^a$ & $\sigma(20$''$)^a$\\
(MHz)      & (mK)          &    (mK)       &  (mK)        \\
\hline
     & P1 P2 P3& P1 P2 P3& P1 P2 P3\\
\hline
31120&3.6 4.7 7.4 & 3.2 3.4 6.0 &2.3 2.6 4.5\\
31500&2.6 3.8 6.1 & 2.3 2.8 4.9 &1.9 2.0 3.6\\
32500&2.2 3.2 6.2 & 2.0 2.9 4.8 &1.4 1.6 3.2\\
33500&2.2 2.8 3.9 & 1.9 2.1 3.7 &1.3 1.6 2.7\\
34500&2.5 3.0 5.0 & 2.2 2.8 4.1 &1.6 1.9 3.2\\
35500&2.5 2.6 4.8 & 2.3 2.2 4.2 &1.7 1.8 2.9\\
36500&2.2 3.2 5.9 & 2.2 2.7 5.0 &1.5 2.0 3.2\\
37500&2.6 3.3 5.8 & 2.4 2.9 4.9 &1.7 1.9 3.2\\
38500&2.7 3.5 6.0 & 2.3 3.3 5.3 &1.6 2.2 3.9\\
39500&2.9 3.5 6.1 & 2.5 3.1 5.2 &2.0 1.9 4.2\\
40500&2.9 4.0 6.3 & 2.6 3.3 4.7 &1.9 2.1 3.9\\
41500&3.4 4.1 7.3 & 3.1 3.6 5.6 &2.1 2.5 4.2\\
42500&3.3 3.8 6.8 & 2.9 3.6 5.6 &2.2 2.5 4.3\\
43500&3.7 5.2 6.1 & 3.2 3.9 5.5 &2.1 2.8 4.5\\
44500&3.2 5.1 6.9 & 2.9 4.3 6.4 &2.2 3.2 4.7\\
45500&3.8 5.6 8.4 & 3.5 4.7 7.4 &2.9 3.4 6.0\\
46500&4.1 6.4 9.1 & 3.9 5.4 7.4 &2.9 3.9 6.3\\
47500&5.5 7.6 10.0& 4.9 6.8 9.0 &3.5 4.8 7.8\\
48500&6.3 7.8 11.6& 5.8 6.9 9.4 &3.7 4.7 7.7\\
49000&6.2 7.1 12.2& 5.5 6.2 10.8&3.9 4.6 8.3\\
49500&6.3 7.1 12.1& 5.7 6.5 10.5&4.0 4.7 7.8\\
\hline                                                                                             
\end{tabular}
\tablefoot{\\
The three selected positions in the map, P1, P2, and P3,
correspond to the central position
$\Delta\alpha$=0$''$,$\Delta\delta$=0$''$, 
to offset $\Delta\alpha$=100$''$,
$\Delta\delta$=0$''$, and to offset $\Delta\alpha$=120$''$, 
$\Delta\delta$=-120$''$, respectively.\\
\tablefoottext{a}{Root mean square noise derived from a polynomial baseline
removal to a frequency range of $\pm$3 MHz (157 channels) around each selected frequency.
The value adopted for $T_{rad}$ is indicated
in brackets. }\\
}
\end{table}

\begin{figure}
\centering
\includegraphics[width=0.47\textwidth,angle=0]{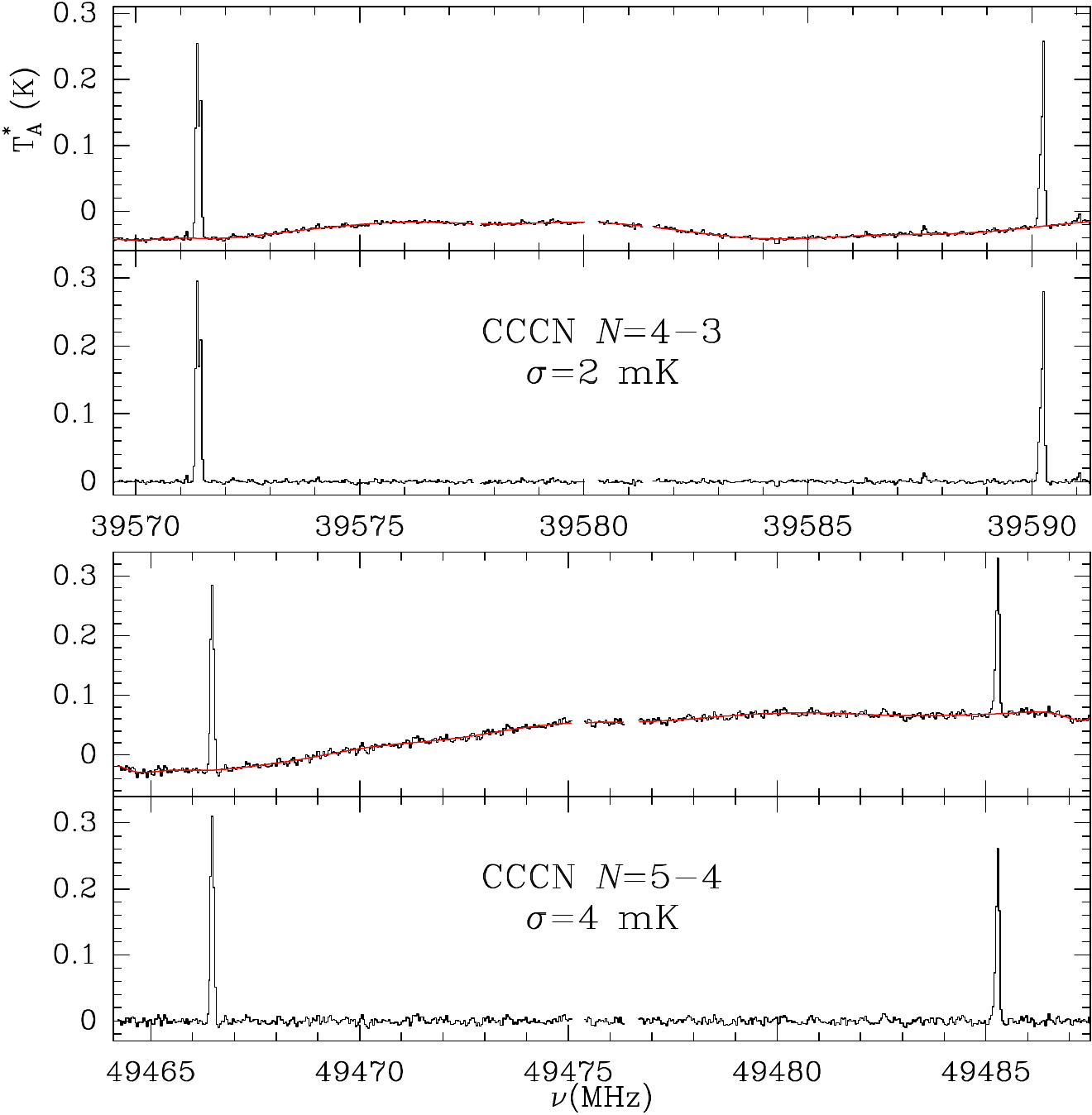}
\caption{Observed lines of CCCN before and after baseline subtraction. For each transition of CCCN
the upper panel shows the raw data and the fitted baseline (red line) at the centre position of the map with $S_{grid}$=20$''$
and $T_{rad}$=20$''$. The panel below each transition shows the data after a polynomial
baseline removal.}
\label{fig:c3n_lines}
\end{figure}

An example of the baseline removal procedure in the  SANCHO maps is shown in Fig. \ref{fig:c3n_lines}
where  we consider the $N$=4-3 and $N$=5-4 transitions of CCCN. These lines
exhibit two line components separated by $\sim$19\,MHz due to the fine structure 
of the rotational transitions.
The lines are shown in Fig. \ref{fig:c3n_lines} before and after baseline subtraction.
For the $N$=4-3 transition the derived sensitivity is 2 mK.
The observed intensity ratio of the two fine components is 1.35, while the theoretical 
value is 1.33 (for optically thin emission). For the two componentes of the $N$=5-4 transition 
the derived sensitivity is 4.2 mK and the line strength ratio is 1.21, which
is very close to the theoretical value of 1.24. 
The number of channels in Fig. \ref{fig:c3n_lines} is 630. 

The folding of the frequency switching data can produce negative features that could
affect the intensity and line profile of some of the observed lines. However, at the level of
sensitivity of SANCHO (2-4 mK) only a few lines with intensities greater than 10 mK
could be affected by this issue. However, at better sensitivities,
such as those of the QUIJOTE line survey \citep{Cernicharo2020,Cernicharo2021a,Cernicharo2023}, 
this effect could be a concern and is mitigated
by the use of two sets of data with different frequency switching throws and
similar sensitivity.

\begin{figure*}
\centering
\includegraphics[width=1\textwidth,angle=0]{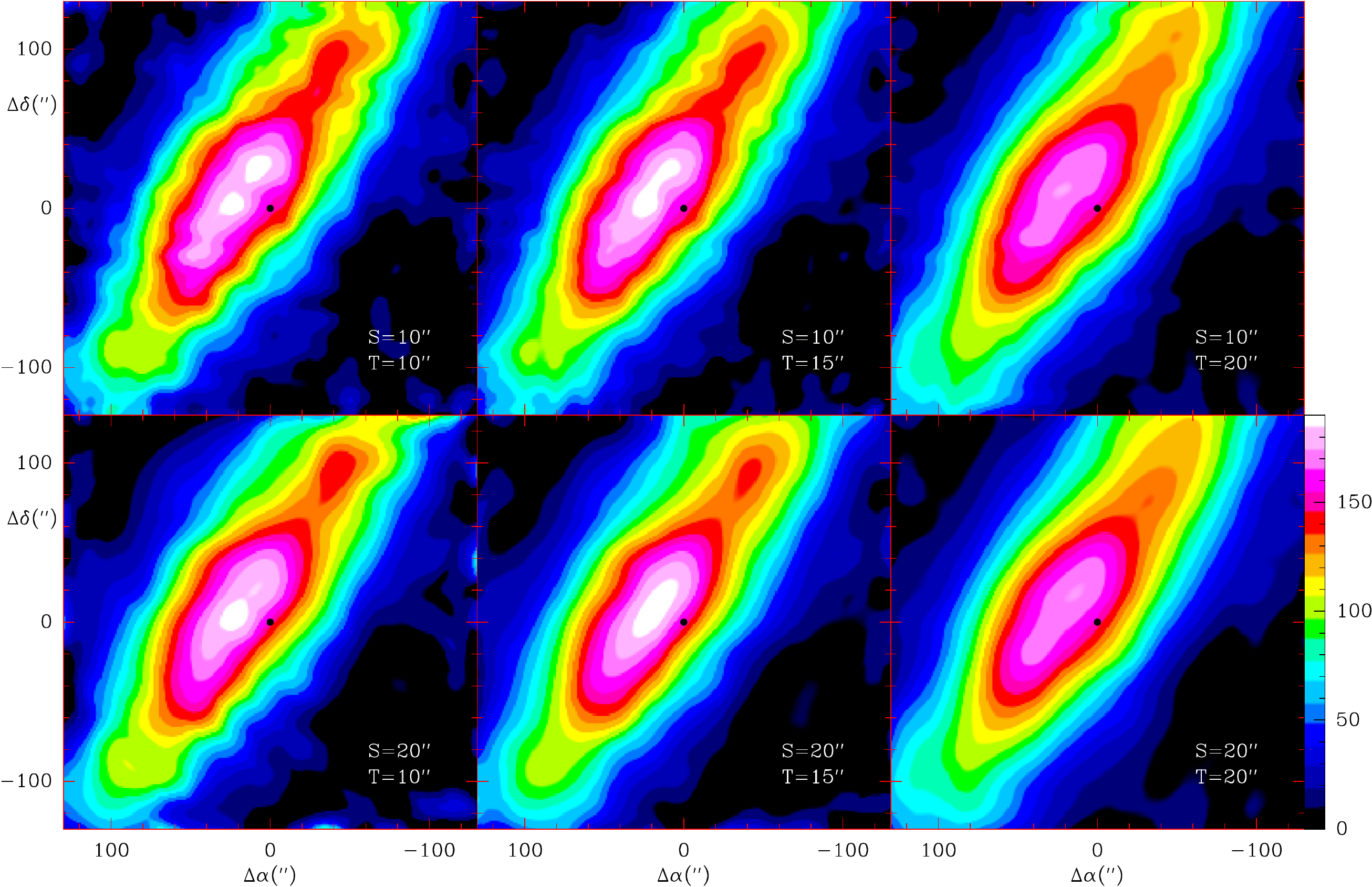}
\includegraphics[width=1\textwidth,angle=0]{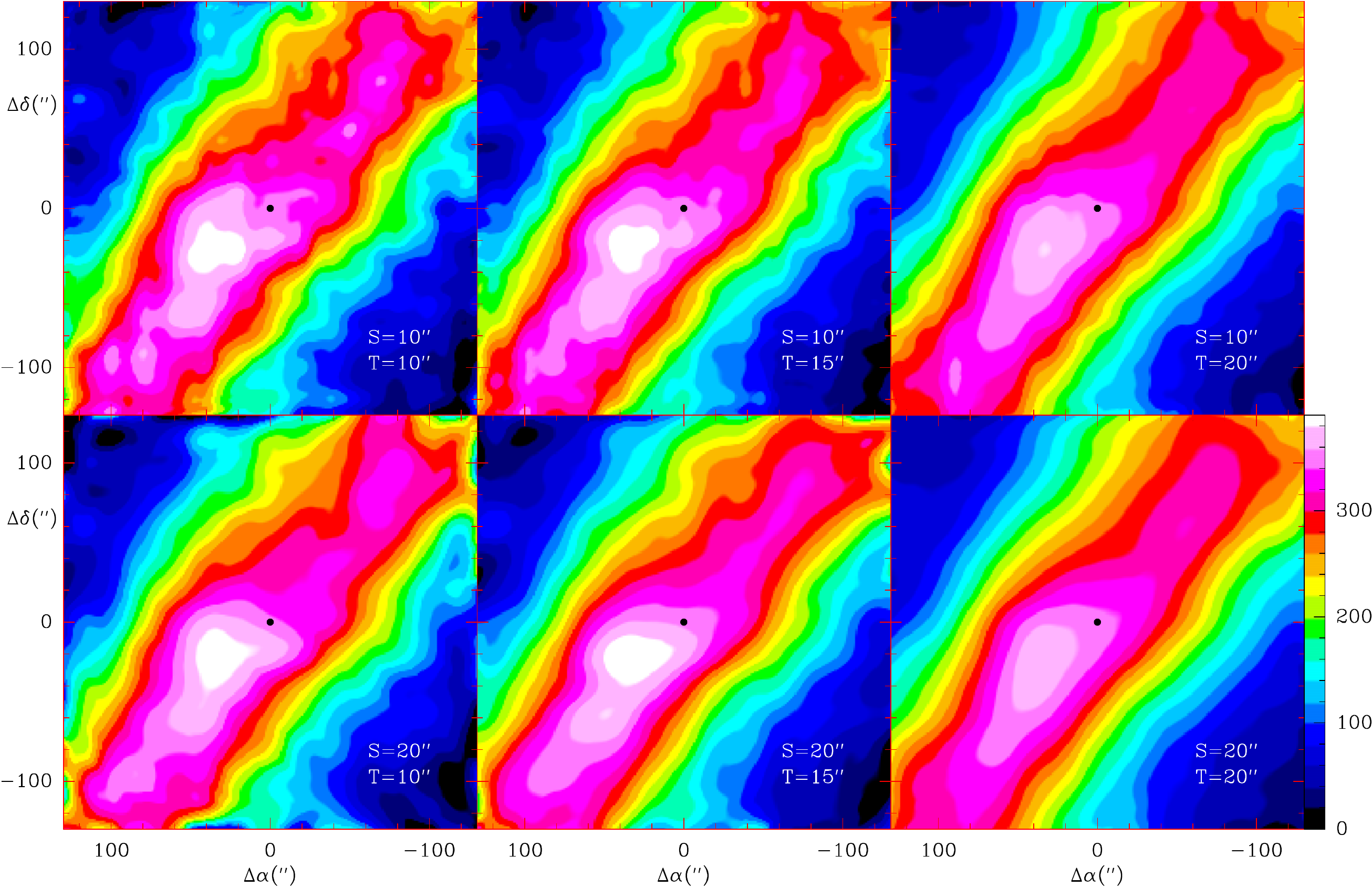}
\caption{Integrated intensity of the $J$=28-27 transition of HC$_7$N (upper panels) and C$^{34}$S $J$=1-0 (lower panels) 
for different values of $S_{grid}$ and $T_{rad}$. The telescope beam size for the two transitions is 56$''$ and 39$''$, respectively.
The black dot indicates the centre of the maps. The maximum value of the integrated intensity changes
by less than 10\% from $T_{rad}$=10" to 20$''$.}
\label{fig:compa_sampling}
\end{figure*}

The effect of the gridding step and tolerance circle on the final aspect of the maps was analysed with
the $J$=28-27 transition of HC$_7$N at 31583.709 MHz (beam size 56$''$), and the $J$=1-0 line of
C$^{34}$S at 48206.942 MHz (beam size 39$''$). Values of $S_{grid}$=10$''$ and 20$''$, and $T_{rad}$=10$''$, 15$''$, and 20$''$ were used.
These maps are shown in Fig. \ref{fig:compa_sampling}.

For HC$_7$N both values of $S_{grid}$ produce oversampled
maps. When $T_{rad}$ is changed from 10$''$ to 20$''$ the maps show a better aspect, but as a counterpart
a considerable spatial smoothing  (smearing)  is introduced, together with a small dilution of the maximum intensity. 
The same applies to the maps of C$^{34}$S, which are
fully sampled in this case. Depending on the physical parameters we want to study, different values
of $S_{grid}$ and $T_{rad}$ can be used depending on the intensity of the mapped lines
and on the desired signal-to-noise ratio in each position on the maps. For weak lines
the spatial shape of the emission could be analysed with $T_{rad}$=20$''$, while the maps for lines with
intensity greater than 100 mK, $T_{rad}$=10$''$ or 15$''$ could be better adapted. In this work we focus
on the study of the extended emission of the weak lines of benzonitrile, and hence we adopt
$T_{rad}$=20$''$ and $S_{grid}$=20$''$.

We   estimated the sensitivity of the adopted values of $S_{grid}$ and $T_{rad}$ by
selecting 21 frequencies across the Q-band. A range of $\pm$3\,MHz (157 channels) 
was selected. Three different positions are considered at distances from the centre of the map
of 0$''$, 100$''$, and 170$''$. The results are given in Table \ref{tab:sensitivity}.

\section{Maps of C$_6$H$_5$CN}
\label{app:stacked}
In order to spectrally stack the lines of C$_6$H$_5$CN and to produce a high signal-to-noise
ratio map of its spatial distribution, we   defined a selection criterion of the
lines to be stacked at each position of the maps.
The intensity of all lines of benzonitrile in TMC-1 in the Q-band are below 10 mK \citep{Cernicharo2021b}. 
Some of these lines have an intensity of 1-2 mK which is well below the detection sensitivity
of the maps with $T_{rad}$=20$''$ (see Table \ref{tab:sensitivity}). In addition, several of these lines show hyperfine structure
\citep[see Fig. B.1 of][]{Cernicharo2021b}.
We   used the last product of the QUIJOTE line survey \citep{Cernicharo2023} to select all lines of benzonitrile that appear as a
single feature and are free of blending from other lines. The QUIJOTE data were also used
to define the windows for baseline removal for each of the selected lines. 

\begin{figure*}
\centering
\includegraphics[width=1\textwidth,angle=0]{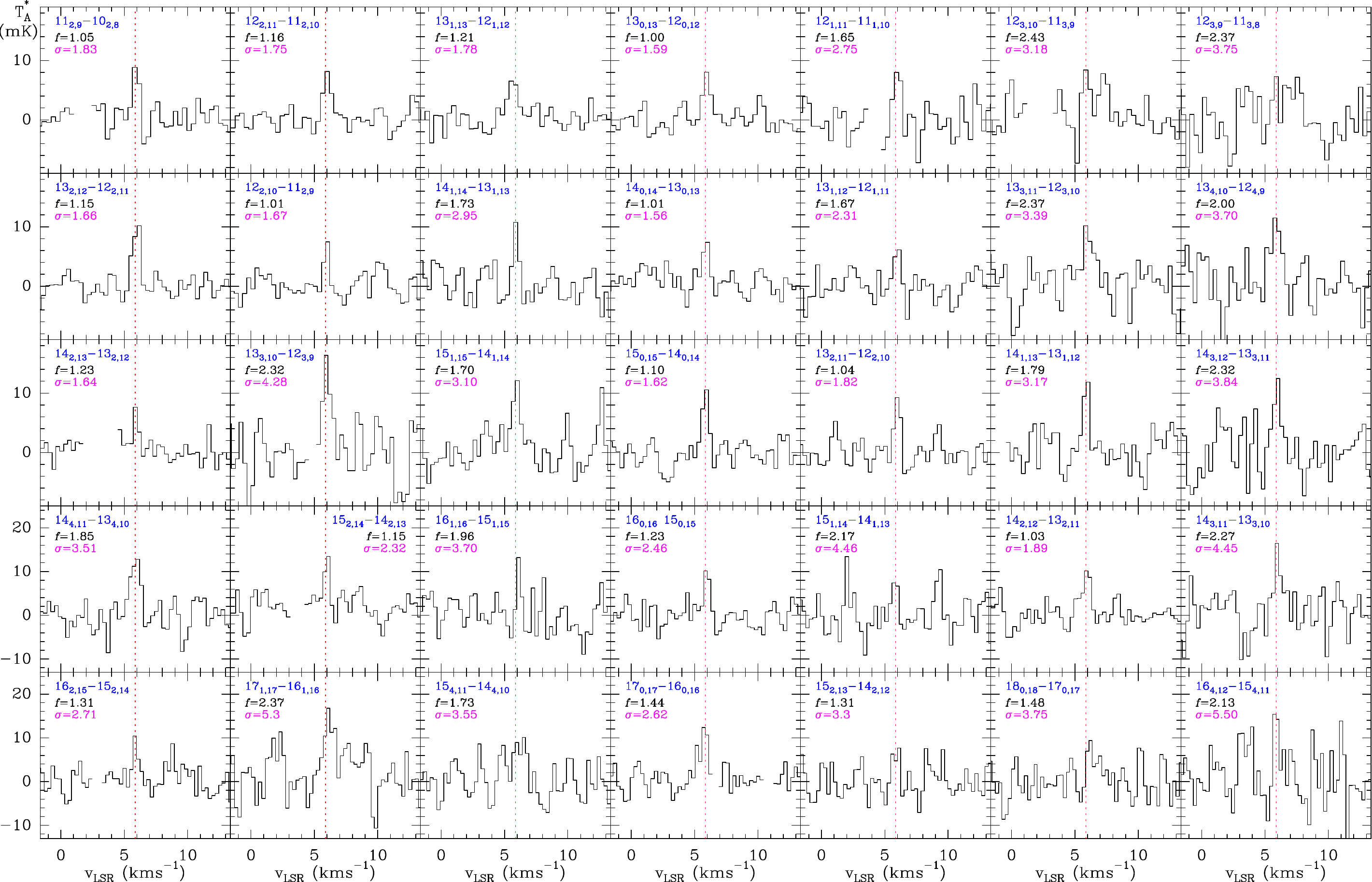}
\caption{Selected transitions of C$_6$H$_5$CN among the 49 lines (see Table \ref{tab:c6h5cn_lines})
used to produce the spectrally stacked map of benzonitrile at each position
of the map. The spectra shown here correspond to the central position. The 
transition quantum numbers are indicated in blue
in the top left   corner of each panel. Each line arises from the individual maps with $S_{grid}$=20$''$ and
$T_{rad}$=20$''$. Each line spectrum was multiplied by the factor $f$ (shown in black in each
panel) to scale its intensity to that of the strongest one (the 13$_{0,13}$-12$_{0,12}$ transition). 
The noise of each spectra, after multiplication by $f$, is shown in violet ($\sigma$; in mK). 
It is used to weight the data during stacking of all the lines at each position of the map (see text). 
The vertical dashed red lines in each panel indicate the local standard of rest velocity of 5.86 km\,s$^{-1}$.
The spectra of the resulting stacked data at different positions
of the map are shown in Fig. \ref{fig:spectra}.}
\label{fig:c6h5cn_00_lines}
\end{figure*}

A second
selection criterion was the intensity of the lines relative to the strongest one that was   taken as reference
(the 13$_{0,13}$-12$_{0,12}$ transition). Only
lines with one-third of the intensity of the reference line in the QUIJOTE data were selected. The final list
of lines used for stacking corresponds to the 49 transitions given in Table \ref{tab:c6h5cn_lines}. This
table also provides the measured intensity of the lines with QUIJOTE \citep{Cernicharo2023} and the applied multiplicative factor for stacking. 
This multiplicative factor scales all lines to the intensity of the reference line.
The data were weighted by 1/$\sigma^2$, where $\sigma$ is the measured
sensitivity of each line after multiplication by the intensity scale factor.
Hence, we can add lines scaled to the same intensity to produce a stacked spectrum 
at each position of the map. 
Most of the selected lines are seen in the central position of the map without stacking, but with a limited signal-to-noise ratio. Figure \ref{fig:c6h5cn_00_lines} shows a sub-sample of the selected transitions at the centre position.
The multiplicative factor and the measured sensitivity (after multiplication to scale the signal to that of the reference
line), are indicated in each panel (see also Table \ref{tab:c6h5cn_lines}). 

The final stacked spectra are shown in Fig. \ref{fig:spectra}.
The improvement of sensitivity after stacking is   a factor of 3.5 with respect to that of the unstacked strongest lines.

\begin{table}
\centering
\caption{Lines of C$_6$H$_5$CN used for spectral stacking at each position.}
\label{tab:c6h5cn_lines}
\begin{tabular}{|c|c|c|c|}
\hline
Transition             & Frequency& T$_A^*$$^a$& $f_{mul}$$^b$\\
                       &   (MHz)  & (mK)   &     \\
\hline
11$_{2,9}$ -10$_{2,8}$ & 32023.223& 9.7    & 1.05\\
12$_{2,11}$-11$_{2,10}$& 32314.461& 8.8    & 1.16\\
13$_{1,13}$-12$_{1,12}$& 32763.506& 8.4    & 1.21\\
13$_{0,13}$-12$_{0,12}$& 32833.827&10.2    & 1.00\\
12$_{1,11}$-11$_{1,10}$& 33249.003& 6.2    & 1.65\\
12$_{3,10}$-11$_{3,9}$ & 33336.175& 4.6    & 2.43\\
12$_{3,9}$ -11$_{3,8}$ & 34389.613& 4.3    & 2.37\\
13$_{2,12}$-12$_{2,11}$& 34864.915& 8.9    & 1.15\\
12$_{2,10}$-11$_{2,9}$ & 34898.780&10.1    & 1.01\\
14$_{1,14}$-13$_{1,13}$& 35202.608& 5.9    & 1.73\\
14$_{0,14}$-13$_{0,13}$& 35249.171&10.1    & 1.01\\
13$_{1,12}$-12$_{1,11}$& 35661.310& 6.1    & 1.67\\
13$_{3,11}$-12$_{3,10}$& 36066.955& 4.3    & 2.37\\
13$_{4,10}$-12$_{4,9}$ & 36321.233& 5.1    & 2.00\\
13$_{4, 9}$-12$_{4,8}$ & 36539.047& 5.0    & 2.04\\
14$_{2,13}$-13$_{2,12}$& 37391.925& 8.3    & 1.23\\
13$_{3,10}$-12$_{3,9}$ & 37477.470& 4.4    & 2.32\\
15$_{1,15}$-14$_{1,14}$& 37638.179& 6.0    & 1.70\\
15$_{0,15}$-14$_{0,14}$& 37668.578& 9.3    & 1.10\\
13$_{2,11}$-12$_{2,10}$& 37709.061& 9.8    & 1.04\\
14$_{1,13}$-13$_{1,12}$& 38038.711& 5.7    & 1.79\\
14$_{3,12}$-13$_{3,11}$& 38772.958& 4.4    & 2.32\\
14$_{4,11}$-13$_{4,10}$& 39147.318& 5.5    & 1.85\\
14$_{4,10}$-13$_{4, 9}$& 39501.269& 4.6    & 2.22\\
15$_{2,14}$-14$_{2,13}$& 39897.901& 8.9    & 1.15\\
16$_{1,16}$-15$_{1,15}$& 40071.306& 5.2    & 1.96\\
16$_{0,16}$-15$_{0,15}$& 40090.922& 8.3    & 1.23\\
15$_{1,14}$-14$_{1,13}$& 40401.006& 4.7    & 2.17\\
14$_{2,12}$-13$_{2,11}$& 40445.137& 9.9    & 1.03\\
14$_{3,11}$-13$_{3,10}$& 40564.446& 4.5    & 2.27\\
15$_{4,12}$-14$_{4,11}$& 41967.092& 5.0    & 2.04\\
16$_{2,15}$-15$_{2,14}$& 42385.689& 7.8    & 1.31\\
17$_{1,17}$-16$_{1,16}$& 42502.779& 4.3    & 2.37\\
15$_{4,11}$-14$_{4,10}$& 42513.064& 5.9    & 1.73\\
17$_{0,17}$-16$_{0,16}$& 42515.315& 7.1    & 1.44\\
16$_{1,15}$-15$_{1,14}$& 42762.895& 4.1    & 2.49\\
15$_{2,13}$-14$_{2,12}$& 43099.531& 7.8    & 1.31\\
15$_{3,12}$-14$_{3,11}$& 43625.762& 4.4    & 2.32\\
16$_{4,13}$-15$_{4,12}$& 44775.824& 4.7    & 2.17\\
18$_{0,18}$-17$_{0,17}$& 44941.103& 6.9    & 1.48\\
17$_{1,16}$-16$_{1,15}$& 45132.664& 4.7    & 2.17\\
16$_{4,12}$-15$_{4,11}$& 45576.805& 4.8    & 2.13\\
16$_{2,14}$-15$_{2,13}$& 45668.515& 6.9    & 1.48\\
17$_{3,15}$-16$_{3,14}$& 46721.270& 3.5    & 2.91\\
18$_{2,17}$-17$_{2,16}$& 47318.725& 4.5    & 2.27\\
19$_{0,19}$-18$_{0,18}$& 47367.825& 5.6    & 1.82\\
18$_{1,17}$-17$_{1,16}$& 47513.403& 3.5    & 2.91\\
17$_{4,14}$-16$_{4,13}$& 47568.715& 3.5    & 2.91\\
17$_{2,15}$-16$_{2,14}$& 48155.082& 5.3    & 1.92\\
\hline                                                                                             
\end{tabular}
\tablefoot{\\
Selected transitions to generate the map of benzonitrile (C$_6$H$_5$CN).
\tablefoottext{a}{Observed intensity in the QUIJOTE line survey in mK.}\\
\tablefoottext{b}{Applied multiplicative factor to the individual map of each
selected transition.}\\
}
\end{table}

\end{appendix}

\begin{thebibliography}{} \tiny
\bibitem[Ag\'undez et al.(2021)]{Agundez2021} Ag\'undez, M., Cabezas, C., Tercero, B., et al. 2021, \aap, 647, L10
\bibitem[Cernicharo(1985)]{Cernicharo1985} Cernicharo, J. 1985, Internal IRAM report (Granada: IRAM)
\bibitem[Cernicharo \& Gu\'elin(1987)]{Cernicharo1987} Cernicharo, J., Gu\'elin, M., 1987, \aap, 183, L10
\bibitem[Cernicharo(2012)]{Cernicharo2012} Cernicharo, J., 2012, in ECLA 2011: Proc. of the European Conference on Laboratory Astrophysics, EAS Publications Series, 2012, Ed.: C. Stehl, C. Joblin, \& L. d'Hendecourt (Cambridge: Cambridge Univ. Press),
251; \texttt{https://nanocosmos.iff.csic.es/?page$\_$id=1619}
\bibitem[Cernicharo et al.(2020)]{Cernicharo2020} Cernicharo, J., Marcelino, N., Ag\'undez, M. et al. 2020, \aap, 642, L8 
\bibitem[Cernicharo et al.(2021a)]{Cernicharo2021a} Cernicharo, J., Ag\'undez, M., Kaiser, R., et al. 2021a, \aap, 652, L9 
\bibitem[Cernicharo et al.(2021b)]{Cernicharo2021b} Cernicharo, J., Ag\'undez, M., Kaiser, R.I. et al. 2021b, \aap, 655, L1 
\bibitem[Cernicharo et al.(2022)]{Cernicharo2022} Cernicharo, J., Fuentetaja, R., Ag\'undez, M., et al. 2022, \aap, 663, L9 
\bibitem[Cernicharo et al.(2023)]{Cernicharo2023} Cernicharo, J., Pardo, J.R., Cabezas, C. et al. 2023, \aap, 670, L19 
\bibitem[Dobashi et al. (2019)]{Dobashi2019}Dobashi, K., Shimoikura, T., Ochiai, T. et al. 2019, \apj, 879:88
\bibitem[Feh\'er et al. (2016)]{Feher2016}Feh\'er, O., T\'oth, L.V., Ward-Thompson, D. et al. 2016, \aap, 590, A75
\bibitem[Foss\'e et al.(2001)]{Fosse2001} Foss\'e, D., Cernicharo, J., Gerin, M., Cox, P. 2001, \apj, 552, 168
\bibitem[Hacar et al. (2013)]{Hacar2013}Hacar, A., Tafalla, M., Kauffmann, J. \& Kov\'acs, A. 2013, \aap, 554, A55
\bibitem[Hacar et al. (2017)]{Hacar2017}Hacar, A., Tafalla, M. \& Alves, J. 2017, \aap, 606, A123
\bibitem[Jones et al.(2011)]{Jones2011} Jones, B. M., Zhang, F., \& Kaiser, R. I. 2011, PNAS, 108, 452
\bibitem[Kirk et al. (2013)]{Kirk2013}Kirk, H., Myers, P.C., Bourke, T.L. et al. 2012, \apj, 766, 115
\bibitem[Lique et al. (2006)]{Lique2006}Lique, F., Cernicharo, J. \& Cox, P. 2006, \apj, 653, 1342
\bibitem[Loison \& Bergeat(2009)]{Loison2009} Loison, J.-C. \& Bergeat, A. 2009, PCCP, 11, 655
\bibitem[Marcelino et al.(2007)]{Marcelino2007} Marcelino, N., Cernicharo, J., Ag\'undez, M. et al. 2007, \apj, 665, L127
\bibitem[McGuire et al.(2018)]{McGuire2018} McGuire, B. A., Burkhardt, A. M., Kalenskii, S., et al. 2018, Science, 359, 202
\bibitem[McGuire et al.(2020)]{McGuire2020} McGuire, B. A., Burkhardt, A. M., Loomis, R. A., et al. 2020, \apj, 900, L10
\bibitem[McGuire et al.(2021)]{McGuire2021} McGuire, B. A., Loomis, R. A., Burkhardt, A. M., et al. 2021, Science, 371, 1265
\bibitem[M\"uller et al.(2005)]{Muller2005} M\"uller, H.S.P., Schl\"oder, F., Stutzki, J., Winnewisser, G. 2005, \jmst, 742, 215
\bibitem[Pardo et al.(2001)]{Pardo2001} Pardo, J.~R., Cernicharo, J., Serabyn, E. 2001, IEEE Trans. Antennas and Propagation, 49, 12
\bibitem[Pety et al.(2005)]{Pety2005} Pety, J., Teyssier, D., Foss\'e, D., et al. 2005, \aap, 435, 885
\bibitem[Pickett et al.(1998)]{Pickett1998} Pickett, H.M., Poynter, R.~L., Cohen, E.~A., et al. 1998, J. Quant. Spectrosc. Radiat. Transfer, 60, 883
\bibitem[Pratap et al. (1997)]{Pratap1997}Pratap, P., Dickens, J.E., Snell, R.L. et al. 1997, \apj, 486, 862
\bibitem[Tercero et al.(2021)]{Tercero2021} Tercero, F., L\'opez-P\'erez, J. A., Gallego, et al. 2021, \aap, 645, A37

\end{thebibliography}
\end{document}